\documentclass[useAMS,usenatbib]{mn2e}
\usepackage{psfig}
\usepackage{graphicx}
\usepackage{graphics}
\usepackage{epsfig}
\usepackage{epsf}
\newcommand{\mnras}{\,{\rm MNRAS}}
\newcommand{\apj}{\,{\rm ApJ}}
\newcommand{\apjl}{\,{\rm ApJL}}
\newcommand{\apjs}{\,{\rm ApJS}}
\newcommand{\aap}{\,{\rm A\&Ap}}
\newcommand{\apss}{\,{\rm Ap\&SS}}
\newcommand{\aj}{\,{\rm AJ}}
\newcommand{\prd}{\,{\rm PRvd}}

\newcommand{\ssst}{\scriptscriptstyle}

\newcommand{\etal}{et al.}

\newcommand{\yr}{\,{\rm yr}}    
\newcommand{\cm}{\,{\rm cm}}

\newcommand{\uG}{\mu{\rm G}}

\newcommand{\no}{n_{\ssst 0}}

\title[Lepto-Hadronic Origin of $\gamma$-rays from the G54.1+0.3 Pulsar Wind Nebula]
{Lepto-Hadronic Origin of $\gamma$-rays from the G54.1+0.3 Pulsar Wind Nebula}

\author[H. Li et al.]
{Hui Li$^{1}$,
Yang Chen$^{1,2}$\thanks{E-mail: ygchen@nju.edu.cn}
~and Li Zhang$^{3}$,\\
$^{1}$Department of Astronomy, Nanjing University, Nanjing 210093, P.\ R.\ China\\
$^{2}$Key Laboratory of Modern Astronomy and Astrophysics, Nanjing University, Ministry of Education, China\\
$^{3}$Department of Physics, Yunnan University, Kunming,P.\ R.\ China\\
}

\begin{document}

\date{Accepted . Received ; in original form}

\pagerange{\pageref{firstpage}--\pageref{lastpage}} \pubyear{2009}

\maketitle

\label{firstpage}

\begin{abstract}
G54.1+0.3 is a Crab-like pulsar wind nebula (PWN) with the highest
$\gamma$-ray to X-ray luminosity ratio among all the nebulae driven
by young rotation-powered pulsars.
We model the spectral evolution of the PWN and find it difficult to
match the observed multi-band data with leptons alone using
reasonable model parameters. In lepton-hadron hybrid model instead,
TeV photons come mainly from $\pi^0$ decay in proton-proton
interaction and the observed photon spectrum can be well reproduced.
The newly discovered infrared loop and molecular cloud in or closely
around the PWN can work as the target for the bombardment of the PWN
protons.

\end{abstract}


\section{INTRODUCTION}\label{sec:intro}
Pulsar wind nebulae (PWNe) are thought to be an efficient accelerator
for cosmic rays with energy above the ``knee". Pulsar, located in
the center of PWN, loses its energy by driving ultra-relativistic
wind of electrons, positrons, and ions. However, it is hard to know
the fraction of energy division of different particle components.
The extended $\gamma$-ray emission from PWN provides an exciting
opportunity for studying the acceleration and radiation mechanism of
particles in ultra-relativistic shocks. It has been long debated
whether the very high energy (VHE) emission from PWNe as well as
from supernova remnants (SNRs) is leptonic or hadronic origin.
Theoretically, it has been suggested that some fraction of the
pulsar's spin-down energy can be converted into nuclei (Cheng et
al.\ 1990; Arons \& Tavani 1994), which indicates that TeV emission
from PWNe may contain contribution from both leptons and hadrons.
Indeed, nucleonic models have been used to reproduce the
$\gamma$-rays from Crab and Vela X, respectively (Atoyan et al.\
1996; Horns et al.\ 2006). Recently, the discovery of TeV emission
from G54.1+0.3 by VERITAS (Acciari et al.\ 2010) presents a brand
new case for highlighting the relative significance of hadrons in
PWNe.

G54.1+0.3 is a Crab-like (Lu et al.\ 2002) SNR with properties very
similar to the Crab Nebula in both morphology and photon spectral
indices. The central pulsar, PSR J1930+1852, has a period of
$P=137$~ms and a period derivative of $\dot P=7.5\times10^{-13} \rm
s~\rm s^{-1}$, corresponding to a current spin-down luminosity of
$L_{sd}=1.2\times10^{37}\rm erg~\rm s^{-1}$ and a characteristic age
$\tau_{c}\approx2900\rm yr$ (Camilo et al.\ 2002). A faint X-ray
shell was most recently detected surrounding the PWN up to $\sim6'$
from the pulsar (Bocchino, Bandiera, \& Gelfand 2009). The SNR has
been suggested to be at a distance of $6.2~\rm kpc$ by the HI
absorption and morphological association with a molecular cloud
(Leahy et al.\ 2008).

Recent AKARI  observation discovered an infrared (IR) loop, which is
explained to be a star-formation loop around the G54.1+0.3 PWN (Koo
et al.\ 2008) and is alternatively explained to be the
freshly-formed dust in the supernova ejecta (Temim \etal\ 2009).
 Using VLA radio polarization and Spitzer
mid-IR observations, Lang et al.\ (2009) found a molecular cloud
located at the southern edge of the PWN and suggested an interaction
between the PWN and the cloud. In $\gamma$-rays, VERITAS observed
the VHE TeV emission from G54.1+0.3 and found that the efficiency of
converting the spin-down energy to $\gamma$-ray emission is high and
the ratio of $\gamma$-ray to X-ray luminosity is as large as 0.7.
This ratio, two orders of magnitudes higher than that of the Crab,
is the highest among all the nebulae supposedly driven by young
rotation-powered pulsars (Acciari \etal\ 2010). This may imply that
the VHE TeV emission has extra components in addition to the
contribution from commonly-acknowledged energetic leptons scattering
background photons. The newly discovered IR loop and/or molecular cloud
around the G54.1+0.3 PWN may act as an appropriate target for the
energetic protons to account for high-efficiency $\gamma$-ray
production from this unusual source (Bartko \& Bednarek 2008).

In this letter,
we show that the TeV emission from G54.1+0.3 cannot be accounted for
by leptons alone, but can be naturally explained by introduction of
a hadronic component.

\section{MODEL AND RESULTS}\label{sec:model}

\subsection{The Pure-Lepton Case}\label{sec:leptonic}

We first try to reproduce the wide-range radiation spectrum of
G54.1+0.3 from radio to TeV using a pure lepton component. For
calculating the spectral evolution of the PWN, we specify the
evolution of the time-dependent injection spectrum and that of the
magnetic field in the following.

Let us consider the relativistic wind of leptons produced within the
light cylinder of the pulsar where the spin-down power L(t) is
injected into PWN. A termination shock is formed in the outflowing
relativistic wind, where the ram pressure is balanced by the
pressure of surrounding medium, and accelerates particles to high
energies. The leptons produced inside the light cylinder of the
pulsar account for the radio emission, while the wind leptons
accelerated by the shock have a Fermi-type energy spectrum and
contribute to the X-ray emission.

As usual, we assume that the injection spectrum of the relativistic
particles $Q_{\rm inj}(\gamma, t)$ obeys a broken power-law
\begin{equation}\label{eq:PWNinj}
Q_{\rm inj}(\gamma, t) = \left\{
\begin{array}{ll}
Q_{\rm 0}(t) (\gamma /\gamma_{\rm {b}})^{-p_{\rm 1}} & \mbox{ for $\gamma_{\rm min} \leq \gamma \leq \gamma_{\rm b}$ ,} \\
Q_{\rm 0}(t) (\gamma / \gamma_{\rm {b}})^{-p_{\rm 2}} & \mbox{ for  $\gamma_{\rm b} \leq \gamma \leq \gamma_{\rm max}$ ,}
\end{array} \right.
\end{equation}
where $Q_{0}$ is normalization coefficient, $\gamma$ is the Lorentz
factor of the relativistic electrons and positrons, and the minimum
($\gamma_{\rm min}$), maximum ($\gamma_{\rm max}$), and break
($\gamma_{\rm b}$) Lorentz factors together with the energy indices
($p_{1}$ and $p_{2}$) are assumed time-independent. Parameter
$\gamma_{\rm max}$ is obtained so as to confine the accelerated
electrons within the PWN (i.e., the electrons's Larmor radius must
be less than the radius of the PWN) (Venter \& de Jager 2006)
\begin{equation}\label{eq:Emax}
\gamma_{\rm max}\approx\frac{e}{2m_{\rm e}c^2}\sqrt{\frac{\sigma L(t)}{(1+\sigma)c}} ,
\end{equation}
where magnetization parameter $\sigma$ is the ratio of the
electromagnetic energy flux to the lepton energy flux at the wind
shock of the PWN. Parameter $\gamma_{\rm min}=100$ is assumed so as
to reproduce the flux of the observed minimum frequency at radio
wavelengths. Bucciantini et al.\ (2010) found that $\gamma_{\rm b}$
is at a similar value in a narrow range of $10^{5}$--$10^{6}$ for
several PWNe of a variety of ages, which is closely related to
the working of pulsar magnetospheres, pair multiplicity, and the
particle acceleration mechanisms. Therefore, here we adopt
$\gamma_{\rm b}=5\times10^{5}$ without loss of generality.

The injection spectrum can be related to the spin-down power $L(t)$
of the pulsar at given time $t$ by assuming that a fraction
($\eta_{e}$) of the spin-down power is converted into lepton
luminosity: $\eta_{e}L(t)=\int Q(\gamma,t)\gamma m_{e}c^{2}d\gamma$.
For a spin-down pulsar, $L(t)=L_{0}[1+(t/\tau_0)]^{-(n+1)/(n-1)}$,
where $L_{0}$ is the initial spin-down power, $\tau_0$ the
characteristic timescale, and $n$ the breaking index (here we adopt
$n=3$ for simplicity).
Thus the normalization parameter $Q_{\rm 0}(t)$ can be derived as
(Tanaka \& Takahara 2010)
{\setlength\arraycolsep{2pt}
\begin{eqnarray}\label{eq:PWNinj0}
&Q_{\rm 0}(t) & =  \frac{L_{\rm 0} \eta_{e}}{m_{\rm e}c^{2}}
 \left(1+\frac{t}{\tau_{\rm 0}}\right )^{-2} \times
\nonumber\\
 & & \left [\frac{\gamma_{\rm b}^{2}(p_{\rm 1}-p_{\rm 2})}
  {(2-p_{\rm 1})(2-p_{\rm 2})} +
 \frac{\gamma_{\rm b}^{p_{\rm 2}}
 \gamma_{\rm max}^{2-p_{\rm 2}}}{2-p_{\rm 2}} -
  \frac{\gamma_{\rm b}^{p_{\rm 1}}
  \gamma_{\rm min}^{2-p_{\rm 1}}}{2-p_{\rm 1}} \right ]^{-1} .
\end{eqnarray}}

On the assumption of magnetic-field energy conservation (see Tanaka
\& Takahara 2010 for the comparison of various approximations of
magnetic field evolution),
\begin{equation}\label{eq:mag-energy}
 \frac{4\pi}{3} R_{\rm{PWN}}^{3}(t) \cdot \frac{B^2(t)}{8\pi}
 = \int_0^{t} \eta_{B} L(t') dt' ,
\end{equation}
the time-varying field strength of the nebula is given by
\begin{equation}\label{eq:mag-evolve}
 B(t)=\left[\frac{6\eta_{B}L_0\tau_0t}{R_{\rm PWN}^3(t+\tau_0)}\right]^{1/2}
\end{equation}
where $\eta_{B}$ is the fraction of spin-down energy converted to
the magnetic energy and $R_{\rm PWN}$ the average radius of the PWN.
(In parenthesis, the magnetization parameter is thus essentially
$\sigma\sim\eta_{\rm B}/\eta_{\rm e}$.)
Because the young G54.1+0.3 PWN ($\sim2900$yr) may be in an
evolution stage before the reverse shock passage (typically at
$1\times 10^4$yr, e.g., Reynolds \& Chevalier; Gelfand 2009), we
also assume that the PWN is freely expanding at velocity $v_{\rm
PWN}$ and thus have $v_{\rm PWN}\sim550(R_{\rm PWN}/1.8 {\rm pc})
(t/2900 {\rm yr})^{-1}$ km s$^{-1}$.

The volume-integrated particle number as a function of energy is
described by the continuity equation in the energy space:
\begin{equation}\label{eq:dist}
 \frac{ \partial}{\partial t} N(\gamma, t) + \frac{ \partial}{ \partial \gamma}
 \left[\dot{\gamma}(\gamma, t) N(\gamma, t) \right] =Q_{\mathrm{inj}}(\gamma, t)-\frac{N(\gamma,t)}{\tau_{esc}(t)}
\end{equation}
where $\dot{\gamma}(\gamma,t)$ is the cooling rates of the
relativistic leptons including the synchrotron radiation, the
inverse Compton scattering off the cosmic microwave background (CMB)
and ambient IR radiation, and the adiabatic expansion, i.e.,
\begin{equation}\label{eq:loss}
\dot{\gamma}(\gamma, t) = \dot{\gamma}_{\mathrm{syn}}(\gamma,t) +
 \dot{\gamma}_{\mathrm{IC}}(\gamma) + \dot{\gamma}_{\mathrm{ad}}(\gamma,t),
\end{equation}
and $\tau_{esc}$ is the escape timescale and can be estimated as in
Bohm diffusion (e.g., Zhang et al. 2008),
\begin{equation}\label{eq:dif}
\tau_{esc}\approx9\times10^{5}\bigg[\frac{B(t)}{80\rm \mu G}\bigg]
\bigg(\frac{E_{e}}{10\rm TeV}\bigg)^{-1}\bigg[\frac{R_{\rm
PWN}(t)}{1.8 \rm pc}\bigg]^{2} \rm yr ,
\end{equation}
where the current magnetic field strength $80\uG$ (see
\S\ref{sec:B}) is used. The adiabatic loss $\dot{\gamma}_{\rm
ad}=-\gamma/t$ is the dominant cooling process for the low energy
particles and insignificant for the high energy ones.

\begin{figure} 
\centerline { {\hfil\hfil
\psfig{figure=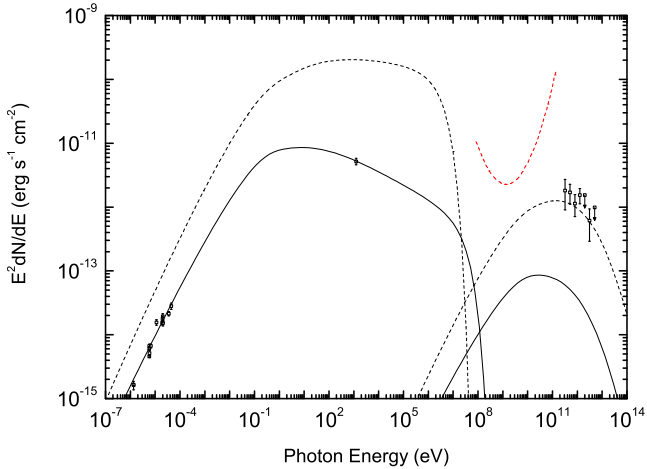,height=2.8in,angle=0}
\hfil\hfil } }
\caption{Comparison of the predicted spectra in the pure-lepton Models A (solid line) and B (dashed line) with the observed data for G54.1+0.3 in
radio (Natasha et al.\ 2008; Lang et al.\ 2009),
 X-rays (Lang et al.\ 2009) and $\gamma$-rays (Acciari et al.\ 2010).
The model parameters are described in the text of \S\ref{sec:leptonic}.
 The red dashed line shows the 1 year, 5$\sigma$ sensitivity
for the {\sl Fermi} LAT ({\sl Fermi} LAT 2007).} \label{fig:syn-IC}
\end{figure}

\begin{figure} 
\centerline { {\hfil\hfil
\psfig{figure=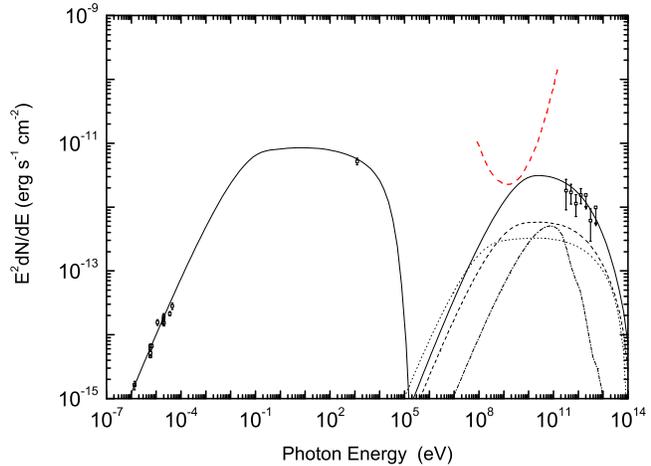,height=2.8in,angle=0}
\hfil\hfil } }
\caption{The same as Figure~\ref{fig:syn-IC}, but for pure-lepton Model C (solid line).
 The parameters are described in the text of \S\ref{sec:leptonic}.
 The IC flux (the solid line on the right side) is dominated by scattering with the IR photons from the SNR, while
 the IC scattering with the IR photons from Galactic diffusion (dashed), the CMB (dotted), and the
 starlight of the possible YSOs (dashed-dotted) are also shown.} \label{fig:both}
\end{figure}

The time-dependent lepton distribution is numerically solved from
the continuity equation (\ref{eq:dist}). Then multi-wavelength
non-thermal emission can be calculated for the process of
synchrotron radiation and inverse Compton scattering, with photon
spectra plotted in Figures~\ref{fig:syn-IC} and~\ref{fig:both} (as
described below).

Here the $\gamma$-rays are considered to purely come from leptons
scattering the soft radiation field (CMB, IR and optical photons in
the Galactic plane, and the IR-optical-UV emission of possible young
stellar objects (YSOs) in the IR loop). The IR background at the
Galactic disc is characterized by temperature $25$K and energy
density two times larger than the CMB, while the optical background
by temperatures between 5000 and $10^4$~K and energy densities equal
to the CMB. The incident IR photons from the SNR are defined by a
$\sim90$~K blackbody radiation with the energy density
$\sim5.3\times10^{-12}\rm erg~cm^{-3}$ based on the Spitzer IRAC
fluxes at $24\mu$m and $70\mu$m from Temim et al.\ (2009), a factor
of roughly 5 larger than the IR energy density in the Crab Nebula
and 13 larger than the energy density in the CMB. In the calculation
we also take into account the possible IR-optical-UV starlight from
11 possible YSOs, which has an energy density
$\sim4.4\times10^{-11}\rm erg~cm^{-3}$ with a blackbody temperature
$T\sim35000K$ (Koo et al.\ 2008).
The IC flux is dominated by scattering with the IR photons from the
SNR, while the IC scattering with other components are insignificant
by comparison. Note that the power of synchrotron self-Compton
emission to synchrotron emission $P_{\rm SSC}/P_{\rm syn}=U_{\rm
syn}/U_{B}<10^{-2}$ (here eq.(27) in Tanaka \& Takahara 2010 is
used), the contribution of $\gamma$-ray emission for G54.1+0.3 PWN
from IC scattering off the synchrotron radiation is negligible.


For the physical parameters of the PWN, we set
$L_{0}\approx1.4\times10^{39}\rm ergs~ s^{-1}$,
$\gamma_{b}=5\times10^{5}$, and $p_{1}=1.2$ according to previous
studies (Camilo et al.\ 2002; Lang et al.\ 2009; Bucciantini et al.\
2010) and leave other three parameters, $\eta_{e}$, $\eta_{B}$, and
$p_{2}$, adjustable. For comparison, we develop three sets of
parameters for leptonic model. In Model A, we reproduce the observed
results of radio to X-ray emission (which are synchrotron) and get
$\eta_{e}=6\%$, $\eta_{B}=8\%$, and $p_{2}=2.4$. As can be seen in
Figure \ref{fig:syn-IC} (the solid line), the resulting TeV emission
from leptons is lower than the observed flux by more than an order
of magnitude. In Model B, we change parameters to reproduce the
observed TeV emission by IC scattering soft photon fields described
above. The adopted parameters are $\eta_{e}=92\%$, $\eta_{B}=8\%$,
and $p_{2}=2.1$. The resulting synchrotron radio and X-ray emission
(the dashed line in Figure \ref{fig:syn-IC}) excess the observation
data by more than an order of magnitude. The current magnetic field
strength for Model A and B, $80\mu\rm G$ (derived from observation,
see \S\ref{sec:B}), has been used in Eq.(\ref{eq:mag-evolve}). In
order to match both the synchrotron and IC emission to the observed
data, we explore the parameter space and obtain the third model
(Model C) (Figure~\ref{fig:both}) with $\eta_{\rm e}=99.8\%$,
$\eta_{\rm B}=0.15\%$, and $p_{2}=2.8$. However, this corresponds to
a weak magnetic field $\sim10\mu \rm G$. If we adopt an age of
$2000\yr$ for this PWN as obtained by Bocchino \etal (2009) in their
dynamic evolution model, other than $2900\yr$, then lower field
strength would be needed in Model C. Such low values of the field
strength are inconsistent with that derived from observation, as
will be discussed in \S\ref{sec:B}.
Therefore, it is hard for a pure-lepton model to reproduce the
radio, X-ray, and TeV data simultaneously, and thus the leptons
alone cannot account for the $\gamma$-ray emission.

\subsection{The Lepton-Hadron Hybrid Case}\label{sec:hadronic}
We now consider the contribution to the TeV emission from a hadronic
component besides the leptonic contribution. In this model, both
leptons and ions extracted from the charged polar cap region are
accelerated in the rotating magnetospheres of neutron stars and PWN
termination shocks (e.g., Zhang \etal\ 2009). For simplicity, we
assume the protons gain energy from central pulsar and are
represented by a power-law spectrum which is common for the
acceleration process. Then the total energy of protons of the PWN is
$W_{p}=\int
A_{p}E_{p}^{-\alpha_{p}}E_{p}dE_{p}=\int_{0}^{t}\eta_{p}L(t')dt'$,
where $\eta_{p}$ is the energy fraction converted to protons,
$A_{\rm p}$ the normalization coefficient and $\alpha_{p}$ the
spectral index of accelerated protons. So the energy released from
the pulsar consists of the kinetic energy of particles ($\eta_{e}$
and $\eta_{p}$) and the magnetic energy ($\eta_{B}$). For the
energy, $E_p$, of the accelerated protons, the rest energy of
protons ($9.4\times10^8$~eV) is adopted as minimum and the energy at
the ``knee" ($3\times10^{15}$~eV) as the maximum. Note that the
energy converted into leptons and magnetic field in Model A is only
a small fraction ($\eta_e=5\%$ and $\eta_B=8\%$, respectively) of
the total spin-down energy of central pulsar. In fact, in the study
of the Vela~X PWN, Horns et al.\ (2006) have questioned where the
remaining energy injected from pulsar is and suggested a hadronic
origin of TeV emission. Hence, we assume $\eta_p=87\%$ in the
lepton-hadron hybrid case (denoted as Model D).

The Bohm diffusion timescale of the PWN particles determined from
Eq.~(\ref{eq:dif}) ($\sim10^{4}\yr$) is much longer than the PWN
age. Therefore, the protons are considered to be well confined in
the PWN and the escape losses of protons are negligible. The cooling
time of p-p interaction is (e.g., Aharonian 2004)
$t_{\rm pp}\approx1.8\times10^{6}(n_{\rm b}/30\cm^{-3})^{-1}\yr$,
much longer than the age of G54.1+0.3, where $n_{\rm b}$ is the
average density of target baryons in the PWN (see below). Hence the
collision losses of the PWN protons are negligible as well. Also
because $t_{\rm pp}$ is almost energy-independent in the energy
region above $1\rm GeV$, the total spectrum of protons remains
unchanged (Aharonian 2004).
The contribution from the secondary leptons that are created by
protons interaction to the overall spectrum is negligible too, as
compared with the dominant contribution of the primary leptons
(Horns et al.\ 2006; Zhang et al.\ 2009).

\begin{figure} 
\centerline { {\hfil\hfil
\psfig{figure=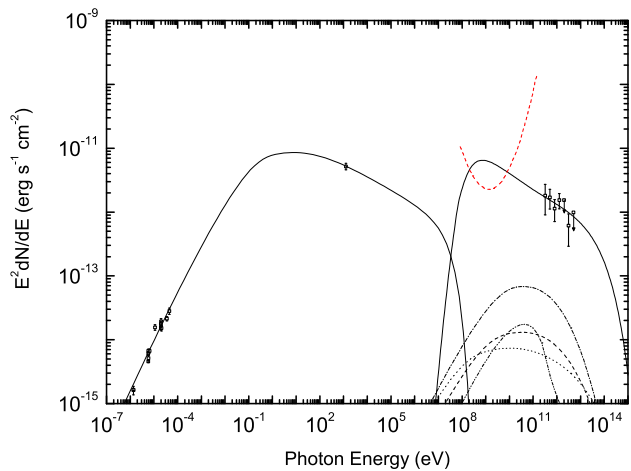,height=2.8in,angle=0}
\hfil\hfil } }
\caption{The same as Fig.\ref{fig:syn-IC}, but for lepton-hadron hybrid Model D. The parameters are described in the test of \S\ref{sec:hadronic}.
The solid line on the right side is dominated by $\pi^0$ decay ensuing from p-p interaction. The inverse Compton scattering with IR photos from SNR
 (dashed-dotted line), IR photos from Galactic diffusion (dashed), starlight of the possible YSOs (dashed-dotted-dotted) and the CMB (dotted) are also shown.} \label{fig:pp-syn}
\end{figure}

In addition to the contribution from the leptons as given in Model
A, we calculate that from p-p interaction so as to match the
observed TeV flux. For the $\pi^0$ decay ensuing from p-p collision,
the analytic emissivity developed by Kelner et al.\ (2006) is used.
It is difficult to determine the detail process of energetic protons
captured by the baryonic targets, since this process depends on
geometry of the PWN and the targets and anisotropy of the magnetic
field and diffusion coefficient. Thus, we assume that a small
fraction ($\xi$) of all hadrons is captured by baryonic targets (as
suggested by Bartko \& Bednarek 2008). The wide-range spectrum of
the PWN can now be well reproduced with
$\xi\sim8\times10^{-3}(n_{\rm b}/30\cm^{-3})^{-1}$ and the results
are shown in Figure~\ref{fig:pp-syn}.
Here a target baryon density $\sim30\cm^{-3}$ has been adopted from
the estimate of the IR clump density (Temin \etal\ 2010); this
number can also be typical of the density of the molecular
materials, which Koo \etal\ (2009) and Lang \etal\ (2010) reported
to detect.  Apparently, even such a low capture efficiency is
sufficient for hadrons to produce the observed flux of TeV emission.



\section{Discussion}\label{sec:discussion}

In the pure-lepton case, Model C seems to marginally match the
wide-range spectrum of the G54.1+0.3 PWN; by comparison, however,
the lepton-hadron hybrid case (Model D) can reproduce the spectrum
better and more physical in the following aspects.

\subsection{Magnetic field}\label{sec:B}
In \S\ref{sec:leptonic}, the field strength obtained in Model C (the
lepton case) is $10\uG$ or even lower.
Such values of field strength are actually weaker than that derived
from observation. Based on radio luminosity, Lang et al.\ (2009)
derived an equipartition field of $38\uG$. However, they suggested
stronger field in the light of the strong polarization which is
organized on large scales of the nebula and implies the PWN is
filled with magnetically-dominated plasma. They also found an
alternative field strength of 80--$200\uG$ by using the lifetime of
the X-ray emitting particles. In Model D (the lepton-hadron case),
however, we use $80\uG$ which can typify the field strength
estimated by Lang et al.


\subsection{TeV index}\label{sec:TeV index}

In Model C, the calculated TeV slope ($\sim2.6$--3) of the IC
spectrum cannot well match the VERITAS data point (with photon index
2.4, Acciari et al.\ 2010). Matching the TeV slope would entail a
lepton ensemble with a unreasonable large energy index 3.8. Even if
the energy losses in high energy leptons are considered, we, using
the time-dependent model, find the energy index of accelerated
leptons by relativistic shock is 2.8, still considerably higher than
the universal power-law index $2.2$--2.3 for Fermi-type acceleration
by the shock of large Lorentz factor using different approaches
(e.g., Horns et al.\ 2007). As a contrast, the observed slope is
easily reproduced by protons p-p interaction with a mild proton
index $\alpha_p=2.4$. This proton index is fortuitously similar to
the lepton index that is used to reproduce the synchrotron X-rays in
Model A.

\subsection{Baryonic targets}\label{sec:environment}
The IR loop closely around the G54.1+0.3 PWN discovered by AKARI was
suggested to be star-forming region (Koo et al.\ 2008), while it was
also argued to be the freshly formed supernova dust heated by
early-type stars belonging to a cluster in which the supernova
exploded (Temim \etal 2009). It was also reported that a molecular
cloud is found to be located at the southern edge of the PWN by the
VLA radio and Spitzer mid-IR observations and thus an interaction
between the PWN with the cloud was suggested (Lang et al.\ 2009).
These components within or surrounding the PWN, whatever they are,
may readily be a baryonic target for the bombardment of the PWN
protons, and therefore it is very reasonable to expect the
$\gamma$-ray contribution from the hadron interaction. This is the
very case that we address in Model D. This scenario seems to
naturally explain the exceptionally high $\gamma$-ray to X-ray
luminosity ratio of G54.1+0.3 among all the rotation-powered PWNe.

The {\sl Fermi} observation at GeV band will be important to
discriminate between the leptonic model and the hadronic model. In
the pure-lepton model (cases A, B, and C; see Figures
\ref{fig:syn-IC} and \ref{fig:both}), the theoretical GeV
$\gamma$-ray flux of the PWN is basically below the 1 year,
$5\sigma$ sensitivity of the {\sl Fermi} LAT, while the
lepton-hadron hybrid model (case D; see Figure \ref{fig:pp-syn})
predicts a GeV flux above the sensitivity.

\section{CONCLUSION}\label{sec:conclusion}

We have calculated the multi-band non-thermal emission from the
G54.1+0.3 PWN in both the pure-lepton case and the lepton-hadron
hybrid case. In the lepton case, we find that the leptons that are
responsible for the radio and X-ray synchrotron cannot alone account
for the TeV $\gamma$-ray emission by IC scattering. An addition of
hadron contribution by p-p interaction can well reproduce the
observation spectrum. The lepton-hadron hybrid scenario is strongly
supported by the most recently discovered IR loop and molecular
cloud in or closely around the PWN. This scenario can also shed
light on the study of the PWNe with high $\gamma$-ray to X-ray
luminosity ratios.




\section*{Acknowledgments}
We thank Q. Daniel Wang, Rino Bandiera, and the anonymous referee
for helpful comments on the manuscripts. Y.C. acknowledges support
from NSFC grant 10725312. L.Z. acknowledges support from NSFC grants
10778702 and 10803005 and Yunnan Province under grant 2009 OC. The
authors also acknowledge support from the 973 Program grant
2009CB824800.

\bsp

\label{lastpage}

\end{document}